\documentclass[aps,amsmath,amssymb,showpacs]{revtex4}

\usepackage{graphicx}
\newcommand{\bml}{\begin{mathletters}}
\newcommand{\eml}{\end{mathletters}}

\newcommand{\f}{\begin{equation}}
\newcommand{\ff}{\end{equation}}
\newcommand{\ul}{\underline}

\begin{document}
%\twocolumn[ \hsize\textwidth\columnwidth\hsize\csname
%@twocolumnfalse\endcsname

\title{Topological Signature of First Order Phase Transitions}

\author{
        L.~Angelani$^{1,4}$,
        L.~Casetti$^{2}$,
        M.~Pettini$^{3}$,
        G.~Ruocco$^{1}$,
        F.~Zamponi$^{1}$.
        }

%\address{
\affiliation{
         $^1$Dipartimento di Fisica and INFM, Universit\`a di Roma
         {\em La Sapienza}, P. A. Moro 2, 00185 Roma, Italy\\
        $^2$INFM - UdR Firenze, via G.\ Sansone 1, 50019 Sesto Fiorentino, Italy\\
        $^3$Istituto Nazionale di Astrofisica, Osservatorio Astrofisico
            di Arcetri, Largo Enrico Fermi 5, I-50125 Firenze, Italy\\
        $^4$ SMC-INFM, Universit\`a di Roma {\em La Sapienza}, P. A. Moro 2, 00185 Roma, Italy}
\date{\today}
%\maketitle
\begin{abstract}
We show that the presence and the location of first order phase
transitions in a thermodynamic system can be deduced by the study
of the {\it topology} of the potential energy function, $V(q)$,
without introducing any thermodynamic measure. In particular, we
present the thermodynamics of an analytically solvable mean-field
model with a $k$-body interaction which -depending on the value of
$k$- displays no transition ($k$=1), second order ($k$=2) or first order
($k$$>$2) phase transition. This rich behavior is quantitatively
retrieved by the investigation of a topological invariant,
the Euler characteristic $\chi(v)$, of some submanifolds of the configuration space. 
Finally, we conjecture a direct link between $\chi(v)$ and the thermodynamic entropy.
\end{abstract}
 %]
\maketitle

Despite their major physical relevance, first-order phase
transitions are still lacking a satisfactory theoretical
understanding of their origin. More specifically, nothing
comparable to the renormalization group analysis for critical
phenomena exists in the case of first order transitions, and, on a
more mathematically rigorous background, neither in the Yang-Lee
theory for the grand-canonical ensemble \cite{YL} nor in the
Ruelle, Sinai, Pirogov theory in the canonical ensemble \cite{RS}
an {\it a priori} mathematical distinction can be made among the
interaction potentials leading to first or second order phase
transitions respectively.

The present work aims at contributing a step forward on this
topic. By simply knowing the microscopic interaction potential,
thus even prior to the choice of a statistical measure, we seek a
 characterization of the occurrence of a phase
transition and of its order. As we shall see in what follows, this
can be actually achieved within the recently proposed topological
approach to the study of Phase Transitions (PT)
\cite{physrep,xy,phi4}. Within this approach, that at
present has dealt only with second order PTs, the existence of a
PT is signaled by a singular point in the energy dependence of
some topological invariant. In particular, in the specific cases
of the mean-field $XY$ \cite{xy} and of the $\varphi^4$ models
\cite{phi4}, it has been shown that it exists a discontinuity in
the first energy derivative of the Euler characteristic,
$\chi(v)$, of some submanifolds of the configuration space,
located at that energy, $v_c$$\equiv$$v(E_c)$, where the system
undergoes a second order PT. The quoted works demonstrated,
therefore, that important thermodynamic features -the second
order PTs- are strictly related to the topology of the potential
energy hyper-surfaces. In particular the knowledge of the
energy distribution of the critical points (saddle-points) of
the potential energy function $V(q)$ allows one to predict the existence
and location of {\it second} order PTs.

The -still lacking- extension of the topological approach to {\it first}
order PTs is obviously of great interest. It would allow to get
new insight into the challenging problem of the origin of first
order PTs, providing a new method of tackling them, and it would
validate the topological approach as a possible theoretical
method of unifying the treatment of different kinds of PTs. The latter point
is particularly interesting in view of encompassing also more
``exotic'' transitions, among which it is worth to mention the
glass transitions. In this respect, we notice that indication on
the existence of a relation between the glass transition and the underlying
topology of the potential energy surface already exist. Indeed, it has
been shown \cite{NOI-cavagna} that, in simulated realistic glasses, the order
 of the saddles visited at a given
temperature $T$ vanishes when $T$ approaches the (dynamical)
transition temperature.

In this work, we show the existence of a relationship between
the topology and the thermodynamics of a dynamical system. In
particular, we introduce and discuss here the "$k$-trigonometric
model", an analytically solvable mean-field model with a $k$-body
interaction. According to the value of the parameter $k$, the
system has no PT ($k$=1), undergoes a second order PT ($k$=2) or a
first order one ($k$$>$$2$). Remarkably, for this model an exact
analytical computation is possible of both the thermodynamics and
of a fundamental topological invariant, the Euler characteristic
$\chi(v)$ of the submanifolds $M_v$ of the configurational
space defined by $V(q)$$\leq$$Nv$. In the
topological properties, the existence of the PT is
signaled, for $k$$>$$1$, in the usual way \cite{xy,phi4}: a discontinuity of the first derivative of $\chi(v)$ is observed at $v_c$$\equiv$$v(E_c)$, i.~e. at
the potential energy
level where the PT actually takes place. Remarkably, {\it this happens also in the case of the first order PT}. But for $k=2$ the second derivative $d^2\chi(v)/dv^2$ is {\it negative} around the singular point, while for $k$$>$$2$
it is {\it positive}.
Therefore: $i)$ A singularity in the first derivative of $\chi(v)$ is a signature of the PT, both in the first or second order case. $ii)$ The sign of the second derivative of $\chi(v)$ around the singularity allows to predict the order of the transition.
The latter conclusion leads to a further, important implication: it is well known \cite{gallavotti} that regions with $d^2S(E)/dE^2>0$ are not thermodynamically stable (they would correspond to
negative specific heat) and their presence corresponds to the existence of a first order PT; therefore, we conjecture a relation between the {\it thermodynamic} entropy and the {\it topological} Euler characteristic.

We study the properties of a specific model, the
mean field $k$-Trigonometric Model ($k$TM), defined by the Hamiltonian
\begin{equation}
H_k(p,q)=K(p)+V_k(q)=\sum_{i=1}^N \frac{p_i^2}{2m} + V_k(q) \ ,
\end{equation}
where the potential energy is given by
\begin{equation}
V_k(q) = \frac{\Delta}{N^{k-1}} \sum_{i_1,...,i_k} [1 -
\cos \frac{2\pi}{L} (q_{i_1}+...+q_{i_k})] \ ;
\label{ptrigm}
\end{equation}
$\Delta$ and $L$ are the energy and length scales. If $k$=1
the $k$TM reduces to the Trigonometric Model (TM) introduced by
Madan and Keyes \cite{MK} as a model for the Potential
Energy Surface (PES) of simple glass-forming liquids that
reproduces many properties of the ``true'' PES of these systems
\cite{NOI-cavagna}. For a given $k$, the model is a mean-field model with
$k$-body interaction. It is convenient to introduce the angular
variables $\varphi_i$$=$$({2\pi}/{L})q_i$. The model has then a
symmetry group $C_{kv}$ obtained by the transformations
$\varphi_i \rightarrow \varphi_i + {2\pi l}/{k}$ and $\varphi_i
\rightarrow -\varphi_i$. If we think at $\varphi_i$ as the angle
between a unitary vector in a plane and the horizontal axis of
this plane, these transformations are rotations in this plane of
an angle ${2\pi l }/{k}$ and  the reflection with respect to the
horizontal axis respectively. Using the relation $\cos (
\varphi_{i_1}+...+\varphi_{i_k})$$=$$\text{Re} ( e^{i\varphi_{i_1}} \cdot
... \cdot e^{i\varphi_{i_k}} ), $ the potential can also be written
as:
\begin{equation}
V_k(\varphi) = N \Delta [ 1 - \text{Re} (c(\varphi) + i s(\varphi))^k ] \ ,
\label{hamil}
\end{equation}
where $c(\varphi)$$=$$\frac{1}{N} \sum_{i} \cos \varphi_i$ and
 $s(\varphi)$$=$$\frac{1}{N}
\sum_{i} \sin \varphi_i$. As usual in simple mean field models it
is easy to calculate the microcanonical partition function, given
by
\begin{equation}
\Omega_{N,k}(E)=\int \frac{d^Np \ d^Nq}{N!} \delta(H_k-E) \ .
\end{equation}
Using the integral representation of the delta function, we get
\begin{equation}
\Omega_{N,k}(E)=\int \frac{d\beta}{2\pi} \int \frac{d^Np \ d^Nq}{N!} e^{-i\beta(H_k-E)} \ .
\end{equation}
Now, as we are looking for a saddle-point evaluation of the integral over $\beta$, we can rotate the integration path on the imaginary axis in the complex-$\beta$ plane. This is justified because, as we will check at the end, the saddle-point is located on this axis \cite{ruota}.
We can now perform the integration over the momenta and use the fact  that $V_k(\varphi)$$=$$V_k(c(\varphi),s(\varphi))$, see Eq. (\ref{hamil}), to obtain
\begin{eqnarray}
\nonumber
\Omega_{N,k}(E)=&{\cal C}_N \ \rho^N \int d\beta \ d\xi \ d\eta \ \beta^{-\frac{N}{2}} \ e^{\beta(E-V_k(\xi,\eta))} \cdot \\ &\cdot \int d^N\varphi \ \delta[N(\xi-c(\varphi))] \ \delta[N(\eta-s(\varphi))] \ ,
\end{eqnarray}
where $\rho$$=$$N/L$ and the constant ${\cal C}_N$ gives only a constant contribution to the entropy per particle, i.\ e. it is at most of order $e^N$.
The last integral can be evaluated using again the integral representation of the delta function, and rotating then the  integration path as previously discussed; it turns out to be:
\begin{gather}
\nonumber
\int \frac{d\mu \ d\nu}{(2\pi)^2} e^{-N(\mu \xi + \nu \eta)} \int d^N\varphi \ e^{\sum_i (\mu \cos \varphi_i + \nu \sin\varphi_i)} = \\
= \int \frac{d\mu \ d\nu}{(2\pi)^2} e^{-N(\mu \xi + \nu \eta)} (2\pi I_0(\Lambda))^N \ ,
\end{gather}
having defined $\Lambda$$=$$\sqrt{\mu^2+\nu^2}$ and the Bessel function
\begin{equation}
I_0(\Lambda)=\frac{1}{2\pi}\int_0^{2\pi}d\varphi \ e^{\Lambda \cos \varphi} \ .
\end{equation}
We can then write the partition function as
\begin{equation}
\Omega_{N,k}(e)= {\cal C}_N \ \rho^N \int d {\mathbf{m}} \ e^{N f_k({\mathbf{m}},e)} \ ,
\end{equation}
where $\mathbf{m}$$\equiv$$(\beta,\xi,\eta,\mu,\nu)$, $e$$=$$E/N$ and
\begin{equation}
\begin{split}
f_k({\mathbf{m}},e)=&\beta e-\beta \Delta [1-\text{Re}(\xi+i\eta)^k] \\ &- \frac{1}{2} \log \beta - \mu \xi -\nu \eta + \log I_0(\Lambda) \ .
\end{split}
\end{equation}
Then, using the saddle-point theorem, the entropy per particle, $s$$=$$S/N$, is given by ($k_B$=1):
\begin{equation}
s_k(e)=\lim_{N \rightarrow \infty} \frac{1}{N} \log
\Omega_{N,k}(e)=\max_{\mathbf{m}} f_k({\mathbf{m}},e) \ .
\end{equation}
To find the maximum of $f_k(\mathbf{m},e)$ one can calculate analytically
some derivatives of $f$ to obtain a one-dimensional problem that can be easily
 solved numerically.
In Fig.~1 we report the caloric curve, i.~e. the temperature vs
energy relation $T(e)$$=$$[{\partial s}/{\partial e}]^{-1}$ for three
values of $k$, $k$=1, 2 and 3. As expected, the temperature is an
analytic function of $e$ for $k$=1, because in this case there is
no interaction between the degrees of freedom. For $k$=2 the
 system undergoes a second order phase transition at a certain
energy value $e_c$, that changes to first order for $k$$>$$2$. At the
transition point the $C_{kv}$ symmetry is broken, the order parameter
being the ``magnetization'' $m$$=$$\langle |c+is| \rangle$; for $e$$<$$e_c$
 there are $k$
pure states related by the symmetry group. It is interesting to
note that -for the present model- this happens also in the case
of the first order transition.

Having an exact solution of the thermodynamics of the system, we can look for some topological signature of the phase transition. In \cite{xy,phi4} it was stated that the Euler characteristic $\chi(v)$ of the submanifolds $M_{v}$$\equiv$$\{q \ | \ V(q) \leq Nv\}$ of the configurational space shows a singularity in correspondence of the potential energy value $v_c$$=$$v(e_c)$ at which the transition takes place. The result of \cite{xy,phi4} was obtained in the case of a second order PT; our aim is to extend this result to the case of a first order PT.
Remarkably, the Euler characteristic of $M_v$ can be calculated {\it analytically}
in our model. The general definition is \cite{nakahara}:
\begin{equation}
\chi(v) \equiv \chi(M_v)=\sum_{n=0}^{N} (-1)^n \mu_n(M_v) \ ,
\end{equation}
where the Morse indexes $\mu_n(M_v)$ correspond to the number of critical points of order $n$ of the function $V(q)$ that belongs to the manifold $M_v$.
The critical points (saddles) $\tilde{\varphi}$ are defined by the
condition $dV_k(\tilde{\varphi})$$=$$0$, and their order $n$ is defined as
the number of negative eigenvalues of the Hessian matrix
${\cal H}^k_{ij}(\tilde{\varphi})$$=$$({\partial^2 V_k}/{\partial \varphi_i
\partial \varphi_j})|_{\tilde{\varphi}}$. 
To determine the location of the saddles we have to solve the system
\begin{eqnarray}
\nonumber
\frac{\partial V_k}{\partial \varphi_j}=-\Delta \ k \ \text{Re}[i(c+is)^{k-1} e^{i \varphi_j}] = \\
\Delta \ k \ \zeta^{k-1} \sin[(k-1)\psi + \varphi_j] = 0 \ , \ \forall j \ ,
\label{saddef}
\end{eqnarray}
where we defined $c$$+$$is$$=$$\zeta e^{i \psi}$. From Eq. (\ref{hamil}) we have $V_k(\varphi)$$=$$N\Delta[1-\zeta^k \cos(k\psi)]$; then the saddles with $\zeta(\tilde{\varphi})$=0 have energy $v$=$V(\tilde{\varphi})/N$=$\Delta$. We can neglect them because, as we will see at the end, $v$=$\Delta$ is a singular point of $\chi(v)$. Then Eq. (\ref{saddef}) becomes
\begin{equation}
\sin [(k-1)\psi + \varphi_j]=0 \ , \ \forall j \ ,
\end{equation}
and its solutions are
\begin{equation}
\label{sadcond}
\tilde{\varphi}_j^{\ul{n}}=[n_j \pi - (k-1) \psi]_{\text{mod} \ 2 \pi} \ ,
\end{equation}
where $n_j$$\in$$\{ 0,1 \}$.
The saddle point $\tilde{\varphi}^{\ul{n}}$ is then characterized by the set $\ul{n}$$\equiv$$\{ n_j \}$.
To determine the unknown constant $\psi$ we have to substitute Eq. (\ref{sadcond}) in the self-consistency equation
\begin{equation} \label{selfcons}
\zeta e^{i\psi}=c+i s=N^{-1} 
%\textbox{\sum_j} 
\sum_j
e^{i\varphi_j}=N^{-1} e^{-i\psi(k-1)} 
\sum_j
 (-1)^{n_j} \ .
\end{equation}
If we introduce the quantity $x(\tilde{\varphi})$ defined by
\begin{equation}
\label{frac_ord}
x=N^{-1}
\sum_j
 n_j \ \Longrightarrow  \ 1-2x=N^{-1} 
\sum_j
 (-1)^{n_j} \ ,
\end{equation}
we have from Eq. (\ref{selfcons})
\begin{eqnarray}
&\label{z-x}
\zeta=|1-2x| \ , \\
&\psi_l=
\begin{cases}
2l\pi /k \hspace{2cm} \text{ for } x < 1/2 \ , \\
(2l+1)\pi /k \hspace{1.12cm} \text{ for } x > 1/2 \ , 
\end{cases}
\label{rhoepsi}
\end{eqnarray}
where $l \in$$\mathbb{Z}$. 
Then the choice of the set $\{n_j\}$ is not sufficient to specify the set $\{\varphi_j\}$ because the constant $\psi$ can assume some different values. This fact is connected with the symmetry structure of the potential energy surface (the different values of $\psi_l$ generate the multiplets of saddles).
We have then obtained that all the saddles of energy $v$$\neq$$\Delta$ have the form
\begin{equation}
\label{sadfin}
\tilde{\varphi}_j^{\ul{n},l}=[n_j \pi - (k-1) \psi_l]_{\text{mod} \ 2 \pi} \ .
\end{equation}
The Hessian matrix is given by
\begin{eqnarray} \nonumber
{\cal H}^k_{ij}=&
\Delta\ k\ Re [N^{-1} (k-1) (c+is)^{k-2} e^{i(\varphi_i+\varphi_j)}  \\
&+\delta_{ij} (c+is)^{k-1} e^{i \varphi_i)} ]  \ .
\end{eqnarray}
In the thermodynamic limit it becomes diagonal 
\begin{equation}
\label{hessdiag}
{\cal H}^k_{ij}=
\delta_{ij}\ \Delta\ k\ \zeta^{k-1}\
\cos \left( \psi(k-1)+\varphi_i \right) \ .
\end{equation}
\begin{figure}[t]
%\epsfysize= 10 truecm
%\begin{center}
\includegraphics[width=.6\textwidth]{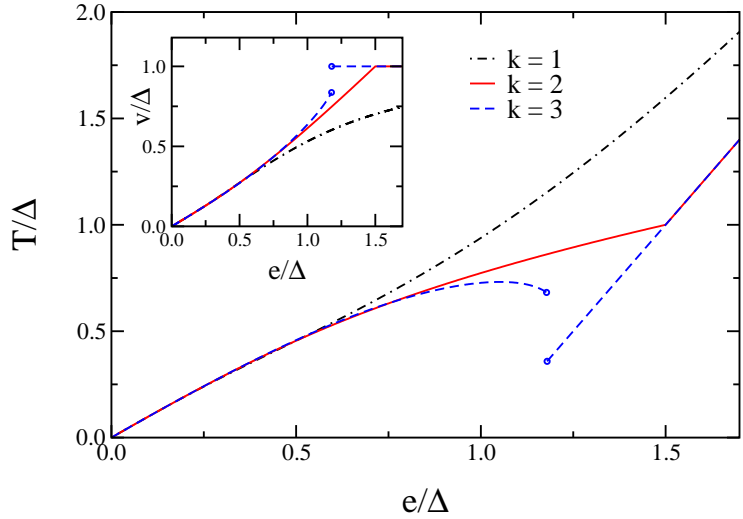}
%\end{center}
\caption{Microcanonical temperature as a function of energy for three different values of $k$; for $k$=1 there is no phase transition, while for $k$=2 there is a second order transition and for $k$$>$$2$ a first order one. In the inset, the potential energy $v$ vs total energy $e$ curve; the critical point is, for $\forall k$$\geq$$2$, $v_c$$=$$\Delta$.}
\label{fig_1}
\end{figure}
One can not {\it a priori} neglect the contribution of the off-diagonal terms to the eigenvalues of ${\cal H}$, but we have numerically checked that their contribution change at most the sign of only one eigenvalue over $N$.
Neglecting the off-diagonal contributions, the eigenvalues of the Hessian calculated in the saddle-point $\tilde{\varphi}$ are obtained substituting Eq. (\ref{sadfin}) in Eq. (\ref{hessdiag}):
\begin{equation}
\lambda_j = (-1)^{n_j} \Delta\ k\ \zeta^{k-1} \ ,
\end{equation}
so the saddle order is simply the number of $n_j$=1 in the set $\ul{n}$; we can identify the quantity $x(\tilde{\varphi})$ given by Eq. (\ref{frac_ord}) with the fractional order $n/N$ of $\tilde{\varphi}$.
Then, from Eq. (\ref{hamil}), (\ref{z-x}) and (\ref{rhoepsi}) we get a relation between the fractional order $x(\tilde{\varphi})$ and the potential energy $v(\tilde{\varphi})$$=$$V(\tilde{\varphi})/N$ at each saddle point $\tilde{\varphi}$:
\begin{equation}
\label{x-v}
x(v)=\frac{1}{2}
\left[1-\text{sgn}\left(1-\frac{v}{\Delta} \right)
\left|1-\frac{v}{\Delta}\right|^{1/k} \right] \ ,
\end{equation}
Moreover, the number of saddles of given order $n$ is simply the number of way in which one can choose $n$ times 1 among the $\{n_j \}$, see Eq. (\ref{sadfin}), multiplied for a constant ${\cal A}_k$ that takes into account the degeneracy introduced by Eq. (\ref{rhoepsi}).
Therefore: $i)$ the fractional
order $x$$=$$n/N$ of the saddles is a well defined
monotonic function of their potential energy $v$, 
given by Eq. (\ref{x-v}), and 
$ii)$ the number of saddles of a given order $n$ is ${\cal A}_k {N \choose n}$.
Then the Morse indexes $\mu_n(M_v)$ of the manifold
$M_v$ are given by ${\cal A}_k {N \choose n}$ if $n/N$$\leq$$x(v)$ and 0
otherwise, and the Euler characteristic is
\begin{equation}
\label{chiv}
\chi(v) ={\cal A}_k \sum_{n=0}^{Nx(v)} (-1)^n  \ {N
\choose n} ={\cal A}_k (-1)^{Nx(v)} {N-1 \choose Nx(v)} \ ,
\end{equation}
using the relation 
$\sum_{n=0}^m (-1)^N {N \choose n}$$=$$(-1)^m {N-1 \choose m}$. 

\begin{figure}[t]
%\epsfysize= 10 truecm
%\begin{center}
\includegraphics[width=.6\textwidth]{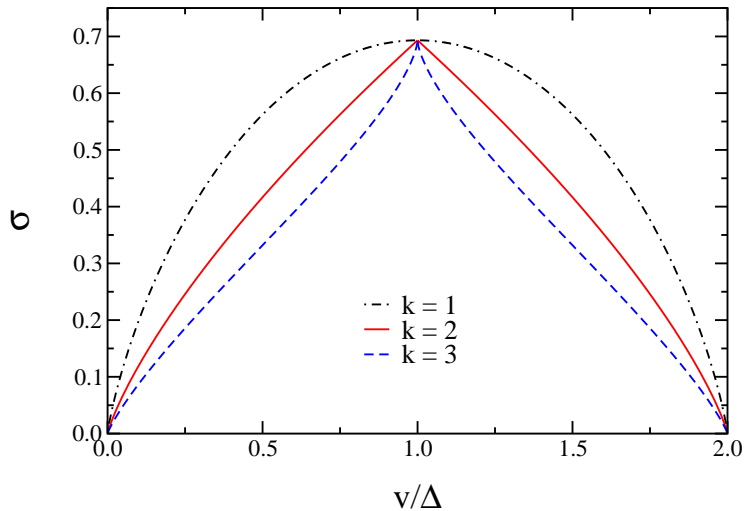}
%\end{center}
\caption{$\sigma(v)$ (see text) as a function of the potential energy. The phase transition is signaled as a singularity of the first derivative at $v_c$=$\Delta$; the sign of the second derivative around the singular point allows to predict the order of the transition. The region $v$$>$$\Delta$, in which $\sigma'(v)$$<$$0$, in not reached by the system (see the inset of Fig. 1).}
\label{fig_2}
\end{figure}

In Fig. 2 we report $\sigma(v)$$=$$\lim_{N\rightarrow \infty}
\frac{1}{N}\log | \chi(v) |$, that, from Eq. \ (\ref{chiv}), is given by:
\begin{equation}
\sigma(v)=  -x(v) \log x(v) -(1-x(v)) \log (1-x(v)) \ .
\end{equation}
It has to be
stressed that $\sigma(v)$ is a purely {\it topological} quantity,
being related only to the properties  of the potential energy
surface defined by $V_k(\varphi)$, and, in particular, to the energy
 distribution of its saddle points.
From Fig. 2 we see that there is a signature of the PT that is evident in 
the analytic properties of $\sigma(v)$. First, we observe that the region $v$$>$$\Delta$ is never reached by the system, as showed in the inset of Fig. 1; this region is characterized by $\sigma'(v)$$<$$0$. We see that: $i)$ for
$k$=1, where there is no phase transition, the function $\sigma(v)$
is analytic;
$ii)$ for $k$=2, when we observe a second order PT, the first derivative of
$\sigma(v)$ is discontinuous at $v_c$=$v(e_c)$=$\Delta$, and its second derivative is {\it negative} around the singular point. $iii)$ for $k$$\geq$$3$ the first derivative of $\sigma(v)$ is also discontinuous at the transition point $v_c$$=$$\Delta$, but its second derivative is {\it positive} around $v_c$. In this case {\it a first order transition takes place}.
Therefore the investigation of the potential energy topology, via $\sigma(v)$, allows us to establish the location and the order of the PTs, without introducing any statistical measure.

The previous results allows us to conjecture that there is a relation between the thermodynamic entropy of the system and $\sigma(v)$. In fact, the presence of a first order transition with a discontinuity in the energy is generally related \cite{gallavotti} to a region of negative specific heat, i.\ e. of positive second derivative of the entropy. Thus, it seems that {\it at least around the transition point} the thermodynamic entropy and $\sigma(v(e))$ are closely related, in the sense that the jump in the second derivative of $s(e)$ is determined by the jump in the second derivative of $\sigma(v(e))$. Then it should be possible to write
\begin{equation}
s(e) \sim \sigma(v(e)) + {\cal R}(e)
\end{equation}
where ${\cal R}(e)$ is analytic (or, at least, $C^2$) around the transition point.
We hope that future work will address this point in a more quantitative way.

In conclusion, we have shown that the presence of a phase transition and its order are signaled in a very clear way in the topology changes of the potential energy hyper-surfaces. In our model, the Euler characteristic of the region of phase space in which $V(q)$$\leq$$Nv$ shows a discontinuity in its first derivative with respect to $v$ {\it if and only if} a phase transition takes place at the same energy value. From the concavity of the Euler characteristic as a function of $v$ around the singular point we can deduce the order of the transition, that is first order if $d^2\chi(v)/dv^2$$>$$0$ and second order if $d^2\chi/dv^2$$<$$0$. From the last observation we conjecture a deep relation between the thermodynamic entropy and some topological property of the potential energy landscape, at least around the transition point.

We acknowledge support from INFM
and MURST COFIN2000.

\end{document}